\documentclass[aps,amsmath,prb,twocolumn,a4paper,showpacs]{revtex4}
\usepackage{graphicx}
\begin{document}

\title{SINGLE WALL NANOTUBES: ATOMIC LIKE BEHAVIOUR AND MICROSCOPIC APPROACH}
\author{S. Bellucci $^1$ and P. Onorato $^1$ $^2$ \\}
\address{
        $^1$INFN, Laboratori Nazionali di Frascati,
        P.O. Box 13, 00044 Frascati, Italy. \\
        $^2$Dipartimento di Scienze Fisiche,
        Universit\`{a} di Roma Tre, Via della Vasca Navale 84,
00146 Roma, Italy}\date{\today}
\date{\today}
\begin{abstract}
Recent experiments about the low temperature behaviour of a Single
Wall Carbon Nanotube (SWCNT) showed typical  Coulomb Blockade (CB)
peaks in the zero bias  conductance and allowed us to investigate the energy levels of interacting electrons.
Other experiments confirmed the theoretical prediction about the crucial role which the long range nature of the Coulomb interaction plays in the correlated electronic transport through a SWCNT with two intramolecular tunneling barriers.

In order to investigate the effects on low dimensional electron systems due to the range of electron electron repulsion, we introduce a model for the
interaction which interpolates well between short and long range regimes.
Our results could be compared with experimental data obtained in SWCNTs and with those obtained for an ideal vertical Quantum Dot (QD).

For a better understanding of some experimental results we also
discuss how defects and doping can break some symmetries of the
bandstructure of a SWCNT.
\end{abstract}


\maketitle

\section{INTRODUCTION}

In the last $20$ years progresses in technology
 allowed the  $construction$ of
several new devices in the range of nanometric dimensions.
The known Moore prediction states that the silicon-data density on a chip doubles every 18 months.
So we are  going toward a new age when the devices in a computer
will live in nanometer
 scale and  will be ruled by the Quantum
Mechanics laws.

QDs\cite{r1,rb}, which could play a central role
within the {\it quantum computation}  as quantum bits (QBIT's)
\cite{BEL,s1,s2,s3,s4} have been studied intensively in the last years\cite{2} thanks to
the advances in  semiconductor technology\cite{tho}.

QDs are small devices, usually formed in semiconductor
heterostructures, with perfectly defined shape and dimensions\cite{2}($<1 \mu m$  in
diameter). They contain from one to a few thousand electrons and,
because of the small volume available, the electron energies
are quantized.
The QDs are useful to study a wide
range of physical phenomena: from atomic like
behaviour\cite{KeM,Tar,OOs} to quantum chaos, to the Quantum Hall
Effect (QHE) when a strong transverse magnetic field acts on the
device \cite{Klprl,Klprb,Klkram,Macdot}.

\
Recently several scientists proposed a new carbon based technology against the usual silicon one. In this sense the discovery of carbon nanotubes (CNs) in 1991\cite{1} opened a new
field of research in the physics at nanoscales\cite{cbmac}.

Nanotubes  are very intriguing systems and many experiments in the
last decade have shown some of their interesting
properties\cite{ebbesen}. An ideal SWCNT is a hexagonal network of carbon atoms (graphene sheet)
that has been rolled up to make a cylinder.
The unique electronic properties of CNs are due to
their diameter and chiral angle (helicity) parametrized by a
roll-up (wrapping) vector $(n, m)$. This vector corresponds to the
periodic boundary conditions\cite{graph,dress} and gives us the
 dispersion relations of the
one-dimensional bands, which link wavevector to energy,
straightforwardly from the  dispersion relation in a graphene
sheet
\begin{equation}\label{eq1}
  \varepsilon_{m,k}=\pm \gamma \sqrt{1-4\cos(\frac{\pi
m}{N_b})\cos(\frac{\sqrt{3}k}{2}) +4{\cos^2(\frac{\sqrt{3}k}{2})}}
\end{equation}
where $N_b$ is the number of periods of the hexagonal lattice
around the compact dimension ($y$) of the
cylinder. 
 If the SWCNT is not excessively doped all the excitations of angular
momentum $m\ne 0$ (corresponding to the transverse motion $k_y=m /
R$)  cost a huge energy of order ($0.3\; \gamma \approx 1$~$eV$), so
we may omit all transport bands except for the lowest one.
From eq.(\ref{eq1}) we obtain  two linearly independent Fermi
points $\pm \vec{K}\approx \pm \frac{2\,\pi }{3\,{\sqrt{3}}}$
 with a right- and a left-moving ($r=R/L=\pm$) branch around
each Fermi point (see Fig.(1.c)).  These branches are highly linear in the  Fermi
velocity $v_F\approx 3 \gamma /2\approx 8\times 10^5$ m/s  up to
energy scales $E < D\approx 1$ eV.
This linear dispersion corresponds to that of a Luttinger Model for  1D electron liquids.
 Many experiments demonstrated a LL
 behaviour\cite{ll,ll2,ll3} in SWCNT\cite{17}with a measurement of  the linear temperature dependence of the
resistance above a crossover temperature $T_c$\cite{Fischer}.

\

In nanometric devices, when the thermal energy $k_B T$ is below the energy for adding an additional electron to the
device ($\mu_N=E(N)-E(N-1)$), 
  low bias (small $V_{sd}$) transport is characterized by a current carried by successive discrete charging and discharging of the dot with  just one electron.

This phenomenon,
known as single electron tunneling (SET) or quantized charge transport, was observed in many experiments in vertical QDs
at very small temperature\cite{KeM,Tar,OOs}.
In this regime the ground
state energy determines strongly the conductance and the period
in  Coulomb Oscillations (COs).
COs correspond to the peaks observed in conductance as a function of gate potential
($V_g$)  and are  crudely described by the CB mechanism\cite{16}:  the  $N-th$ conductance peak occurs when
\cite{Klkram}
$\alpha e V_g(N)=\mu_N$  where
$\alpha=\frac{C_g}{C_\Sigma}$ is the ratio of the gate capacitance
to the total capacitance of the device.
The peaks and their shape  strongly depend on the temperature as
explained by the Beenakker formula for the resonant tunneling
 conductance\cite{16,r1}
\begin{equation}\label{gt}
  G(V_g)=G_0\sum_{q = 1}^{\infty }\frac{V_g -
{\mu_q}}{{k_B}\,T\,\sinh (\frac{V_g - {\mu_q}}{{k_B}\,T})}
\end{equation}
here $ \mu_1,...,\mu_N$ represent the positions of the peaks.

Many experiments showed the peaks in the conductance of QDs and a famous one\cite{Tar} also
showed atomic-like properties of a vertical QD. There
the {\it "addition energy"}, needed to place an extra
electron in a semiconductor QD, was defined  analogously to the
electron affinity for a real atom and was  extracted  from measurements.
 The typical addition energy  ranges from $10$ to
$20 meV$, while
 the disappearance of the COs happens above $\approx50- K$.

\

The transport in CNs often differs from the quantum CB
theory for QDs, because of  the one-dimensional nature of the correlated electrons: so we could need a peculiar theory for resonant tunneling in LLs\cite{812,NG}.
More recently a novel tunneling mechanism\cite{cuniberti} was introduced i.e.  correlated sequential tunneling (CST), originating  from the finite range nature of the Coulomb interaction in SWNTs, in order to  replace conventional uncorrelated sequential tunneling. It
dominates resonant transport at low temperatures and
strong interactions and its prediction  agree with experiments\cite{postma}.
In the experiment a short nanotube segment was created with an addition energy larger than the thermal energy at room temperature ($T_R$), so that the SET can be observed also at $T_R$. The conductance was observed to follow a clear power law dependence with decreasing temperature, pointing at a LL behaviour in agreement with Ref.\cite{cuniberti}.

An  interesting observation about the transport in CNs concerns  the long-range nature of the
Coulomb interaction, which  induces dipole-dipole correlations between the
tunneling events across the left and right barrier\cite{cuniberti}.
The crucial question of the range of the interaction in CNs was investigated in different transport regimes e.g.
in order to explain the LL behaviour of large Multi Wall\cite{noi1} and doped\cite{noi2} CNs.

\

However many experiments showed COs:
e. g. in 1997 Bockrath and coworkers\cite{11}  in a rope of CNs below about $10 -K$ observed
dramatic peaks in the conductance as a function of the
gate voltage 
according the theory  of single-electron charging and resonant tunneling through
 the quantized energy levels of the nanotubes composing the rope\cite{11}.
In this regime
also a SWCNT behaves as an artificial atom  and reveals its shell
structure\cite{10}(the data were taken at $5 mK$).
Recent measurements report
 clean "closed" nanotube dots showing complete
CB\cite{cobden}, which  enable us to
deduce some properties from the addition energy of SWCNT and discuss the role which the Coulomb interaction could play in a 1D system at small
temperatures ($T=0.1\div 0.3-K$).

\

In this paper we analyze the effects of a long range electron
electron interaction in order to determine the addition energy
for models chosen by a vertical QD and a SWCNT in the
Hartree Fock (HF) approximation.

In section II 
we introduce microscopic models and corresponding Hamiltonians  for SWCNTs and QDs also
focusing on a theoretical model for the interaction potential which interpolates between short and long range type interaction.
In section III we show our results about the SWCNT  and QDs
and discuss the effects of asymmetries experimentally
observed\cite{cobden}.

\section{Microscopic approach}

As showed in eq.(\ref{eq1}) near each Fermi point we obtain that
the CN could be  represented by a typical Luttinger Hamiltonian
with linear branches depending on $\overline{k}=k +\alpha K_F$
($\alpha=\pm 1$ labels the Fermi point). We introduce  the
operators that create the  electrons near one of the  Fermi points
(label $\alpha$) belonging to one of the two branches (label
$\zeta$ corresponding to the sign of  $k$),
$\widehat{c}^{}_{\alpha,\overline{k},s}$ and
$\widehat{c}^{\dag}_{\alpha,\overline{k},s}$.
In terms of these operators  the free and interaction Hamiltonians
can be written as
\begin{eqnarray}\label{h0}
  H_0&=&v_F\sum_{\alpha,\overline{k},s}|\overline{k}|
c^{\dag}_{\alpha,\overline{k},s} c_{\alpha,\overline{k},s} \\
\label{hi}
 H_{int}&=&  \sum_{\{\alpha_i\}}\sum_{k,k',q,s,s'}\left(
V_{k,p,s,s'}^{\{\alpha_i\}}(q) c^{\dag}_{\alpha_1,k+q,s}
c^{\dag}_{\alpha_2,p-q,s'}
c_{\alpha_3,p,s'}c_{\alpha_4,k,s}\right) .
\end{eqnarray}

The interaction $V^{\{\alpha_i\}}_{\zeta,\zeta'}(q)$ plays a
central role in order to determine the properties of the electron
liquid.

\

A very crucial question is the effective range of the potential
and its possible screening in a CN. If we denote by $x$ the
longitudinal direction of the tube and $y$ the wrapped one, the
single particle wave function for  each electron reads
$$\varphi_{\zeta,\alpha}(x)=u_{\zeta,\alpha}(x,y)\frac{e^{i \alpha K_F x}e^{\zeta i k x}}{\sqrt{2 \pi L}}$$
where $u_{\zeta,\alpha}(x,y)$ is the appropriate linear combination of
the sublattice states $p=\pm$ and $L$ the length
of the tube. So we can obtain a simple 1D interaction potential
as follows:
\begin{eqnarray}\label{intpot}
\nonumber U^{\zeta,\zeta'}_{\{\alpha_i\}}(x-x') &= &\int_0^{2\pi
R} dy dy' \; u^{\ast}_{\zeta\alpha_1}(x,y)
u^{\ast}_{\zeta'\alpha_2}(x',y')
\\ \nonumber \times \; U^{\zeta,\zeta'}_0(x&-&x',y-y')\;
u^{}_{\zeta'\alpha_3}(x',y') u^{}_{\zeta\alpha_4}(x,y) \;.
\end{eqnarray}
These potentials only depend on $x-x'$ and  the 1D fermion
quantum numbers while  $U^{\zeta,\zeta'}_0(x-x',y-y')$ is obtained
from a linear combination of  $U(x-x',y-y'+p d \delta_{p,-p'})$
sublattice interactions\cite{eg}.

Because of the screening of the interaction in CNs and the
divergence due to the long range Coulomb interaction in 1D
electron systems, it is customary to introduce models, in order to
describe the electron electron repulsion. The usual model is the
so called Luttinger model, where the electron electron repulsion
is assumed to be a constant in the space of momenta, corresponding to
a very short range 1D potential (Dirac delta).
 In order to analyze
the effects of long or short range interactions, we introduce a
model for the electron electron potential depending on a parameter
$r$, which measures the range of a non singular interaction; it
has as limits the very short range potential ($r\rightarrow
0$, delta function) and the infinite long range one ($r\rightarrow
\infty$, constant interaction). So we can conclude that our general
interaction model ranges from the very short range one to the
infinity long range one and eliminates the divergence of the
Coulomb repulsion. In this sense we suppose that our model is good
for describing the interaction, if we do not take in account the IR
and the UV divergences.
\begin{equation}\label{intx}
  U_r(|x-x'|)=U_0\left(\frac{e^{-{\frac{|x-x'|}{r}}}}{2
r}+\frac{r^2}{ r^2+|x-x'|^2}\right)\;.
\end{equation}
  The interaction
between two different electrons with momenta $k$ and $q$ follows
  from the previous formula
$$
V(p=|k-q|)=\frac{1}{{L}^2}\int_0^L dx \int_{-x}^{L-x} e^{i p
y}U(y)dy  \;.
$$
In the limit $L\rightarrow \infty$ we can calculate
the Fourier transform of eq.(\ref{intx})
\begin{equation}\label{intk}
 V_r(|q-q'|)=V_0\left({ \pi r} e^{-{r|q-q'|}}+\frac{1}{ 1+
r^2|q-q'|^2}\right)  \;.
\end{equation}
The scattering processes are usually classified according to the
different  electrons involved and the coupling strengths $g$ are
often taken as constants. This assumption corresponds to the
usual Luttinger model, so we follow this historical  scheme, in
order to classify the interactions.
 The backscattering $g_1^{s,s'}$ involves
electrons in  opposite branches with a large momentum transfer
($q\approx 2 k_F$) so $g_1^{s,s'}\approx V(2 k_F)$. The forward
scattering occurs between electrons in opposite  branches $g_2$
with a small momentum transfer ($q \ll 2 k_F$) so
$g_2^{s,s'}\approx V(0)$ (or $g_2^{s,s'}\approx V(2\pi/L)$). The
forward scattering in the same branch ($g_4$)  involves the pairs
$(k\approx k_F ,p\approx k_F)$ and gives $g_4^\parallel \approx
V(0)-V(p-k)$ and
 $g_4^\perp \approx V(0)$.
Further below, where we discuss the electron electron
interaction, we recall the values of the $g$ constants.

 \

Now we want to introduce the analogous model for a semiconductor ideal
QD. Usually we describe the dot like a 2D system with an
harmonic confinement potential $V(r)=\frac{1}{2}m^*{\omega_d}^2
r^2$ according to measurements\cite{Tar} that demonstrated
that vertical QDs have the shape of a
disk where  the lateral confining potential has a cylindrical
symmetry with a rather soft boundary profile.
Under this hypothesis the quantum single particle levels
depend just on $n=n_++n_-$ (the angular momentum is $m=n_+-n_-$)
$$\varepsilon_n=\hbar \omega_d(n+\frac{1}{2}).$$
The symmetry leads to sets of degenerate single-particle states
which form a shell structure: each shell ($\varepsilon_n$)  has
$2(n+1)$ degenerate states so that the shells  are completely
filled for $N = 2, 6, 12, 20, etc.$  electrons in the dot (Magic
Numbers).

The many body Hamiltonian corresponding to eq.(\ref{h0}) and
eq.(\ref{hi}) 
 has the form
\begin{eqnarray}\label{h0d}
 \hat{H}=\sum_\alpha^\infty \varepsilon_\alpha \hat{n}_\alpha+
\frac{1}{2} \sum_{\alpha,\beta,\gamma,\delta}
V_{\alpha,\beta,\gamma,\delta}\,\,
\hat{c}^\dag_{\alpha}\hat{c}^\dag_{\beta}\hat{c}_{\delta}\hat{c}_{\gamma}.
\end{eqnarray}
Here  $\alpha\equiv(n,m,s)$ denotes the single particle state in
the single particle energy level $\varepsilon_\alpha$,
 ${\hat{c}}^\dag_{\alpha}$ creates a particle in the state $\alpha$ and
$\hat{n}_\alpha\equiv \hat{c}^\dag_{\alpha}\hat{c}_{\alpha}$ is
the occupation number operator. In the following sections we
discuss the essential question regarding the electron electron
interaction in the dot ($V^{n,n'}_{m,m'}$) and analyze in detail the
screening of the effective potential.

\section{Single Wall Carbon Nanotube: Low Temperature Behaviour and Coulomb Blockade}

Before proceeding with the  calculations, we want to point out  that the
real band structures of measured CNs show some differences with
respect to the  ideal case discussed (eq.(\ref{h0})): in order to clarify this point  we
shortly discuss the model and the results of two recent experiments.

To begin with,  we have to introduce a quantization due to the finite longitudinal size of the
tube ($L$) in the dispersion relation  eq.(\ref{eq1}). The
longitudinal quantization introduces a parameter which also gives a
thermal limit for the Atomic Like behaviour: in fact $k$
wavevectors have to be taken as a continuum if $K_B T$ is as a
critical value $E_c=v_F\frac{h}{L}$ and as a discrete set if the
temperature is below (or near) $E_c$.

After the quantization we obtain shells with an 8-fold degeneracy
due to $\sigma$ (spin symmetry), $\alpha$ ($K,-K$ lattice
symmetry), $\zeta$ ($(k-K),(K-k)$ Luttinger symmetry).

Recent experiments
do not support such a high symmetry  and  different
hypotheses were formulated in order to explain this discrepancy.

According to Cobden and Nygard\cite{cobden}  {\it
"the sole orbital symmetry is a two-fold one,
 corresponding to a K-K' subband degeneracy and resulting
 from the equivalence of the two atoms in the primitive cell of graphene structure"}.
Experimentally one can answer this question by observing the
grouping of the peaks in plots of the conductance versus the gate potential.
However in the experiment no four-fold grouping  was observed
because degeneracy was lifted by a mixing between states due
either to defects or to the contacts.

A different  experiment\cite{liang} displays conductance peaks in
clusters of four, indicating that there is a four fold
degeneracy. In ref.\cite{liang} two different shell filling models
are put forward: the first one, when the subband mismatch
dominates, predicts that the spin in the SWCNT oscillates between
$S=0$ and $S=1/2$.

\

In order to take into account the strong asymmetries measured experimentally we
modify the dispersion relation. A first correction has to be
introduced because of the {\em "longitudinal incommensurability"}:
in general $K$ is not a multiple of $\pi/L$ so $K=(N+\delta
N)\frac{\pi}{L}$ with $\delta N<1$ and the energy shift is $\Delta
\varepsilon= v_F \frac{h\delta N}{L}$. A second correction is due
to the subband mismatch ($\delta_{SM}$). The single electron
energy levels are
\begin{equation}\label{ep9}
  \varepsilon_{l,\sigma,p}=\hbar v_F |l\frac{\pi}{L} + p
K|+\frac{(1-p)}{2}\delta_{SM}
\end{equation}
where $p=\pm1$.%
\begin{figure}
\includegraphics*[width=0.9\linewidth]{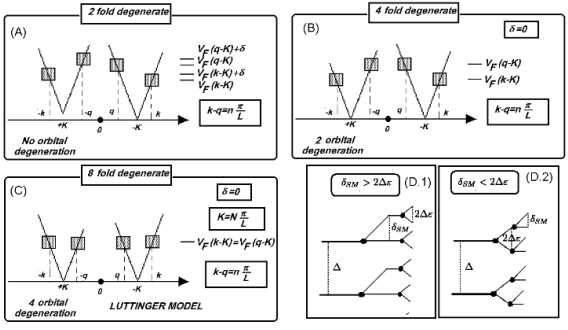}
 \caption{{The dispersion relation and the quantized
levels. The boxes in figure represent energy levels and can be
filled by a pair of electrons with opposite spins. a) The general
case  without any degeneracy. b) The $4$-fold degeneracy case with
$\delta_{SM}=0$. c) The $8$-fold  degeneracy case. d) Differences in
the splitting due to the comparison between $\delta_{SM}$ and
$\Delta \varepsilon$.
}}
\end{figure}

Each choice of parameters gives a different degeneracy for the
quantum levels: the $8$-fold  degeneracy   corresponds to
$\delta_{SM}=0$ and $K=n \frac{\pi}{L}$; the $4$-fold degeneracy is
found if we put just $\delta_{SM}=0$ and the  2 fold degeneracy represents the
general case (see Fig.(1)).

\subsection{High band structure symmetry: damping in the addition energy oscillations}

In the discussion which follows we analyze the effects of the
range of the Coulomb interaction in a simplified system with just two
linear symmetric branches. In this model each shell is filled by
$4$ electrons with opposite momenta and spin.
We can
look for the conservation laws of our Hamiltonian and find the
Number of electrons ($N$),  the Energy, the  total linear momentum
$K=0$ and the spin $S,S_z=0$.

The Fermi sea corresponds to the state where all  the shells with
energy below the Fermi energy ($E_F= v_F n_F h/L$)  are totally
filled ($N_F=4 n_F$). This state will be our ground state
$\Psi_0(N_F)$ and this is true also for interacting electrons in
absence of correlation (i.e. in the HF approximation ).

{The effects of correlation will be discussed in a further
article, here we have to explain the range of validity of our
approximation. The first thing to consider is the interaction
strength $g \approx V(0)-V(2k_F)$ compared to the kinetic energy
$v_F h /L$. The HF approximation is valid if $g\ll v_F h /L$.
However also the temperature plays a central role, in fact if the
temperature increases we have to take into account more excited
states (a sort of thermal cut-off corresponds to  the energy $k_B
T$) so if we are at a very low temperature we can assume the Fermi
sea state as the ground state. }

\

At this point we are able to calculate the \textit{addition energy
($E^A_N$)} following the \textit{Aufbau} sequence explained in
Appendix A.
 $E^A_N$ has a
maximum for some numbers
 $4,8,12,.. 4 n$ (n integer)
due to the shell filling. The shells are filled sequentially and
 { Hund's rule} determines whether a spin-down or a spin-up electron is added so
that the singlet ($S=0$) energy for a $4n+2$ system is always
greater than the triplet ($S=1$) one. Obviously this is an effect
of interaction and is quite different for the long and short range
models.

As we show in Fig.(2) the oscillations due to  Hund's rule
correspond  to the short range potential while the attenuation of
these oscillations when the number of electrons in the 1D system
increases is due to the long range interaction. So we can
draw the following conclusions:

 $\diamond$ The $4-$fold
degenerate  model predicts oscillations in the addition energy due
to the Hund's rule quite similar to the ones
observed in QDs.

$\diamond$ The oscillations periodicity is 4 for this model (8 for
a system with two Fermi points)

 $\diamond$ The
oscillations amplitude  is due to an exchange
 term (proportional to the short range interaction).

$\diamond$ The effect of a long range interaction is a damping of
the oscillations when the number of electrons in the system
increases.

The model with  8-fold degeneracy (see Fig.(1)) has two basic
symmetries: $k \rightarrow -k$, $K L/\pi=N_K$ and usually the
interaction between the electrons with momenta near $K$ and the
ones with momenta near $-K$ is very small so that we have two
independent 4-fold degenerate  Hamiltonians ($n_F=N/4<<N_K$).

\begin{figure}
\includegraphics*[width=0.9\linewidth]{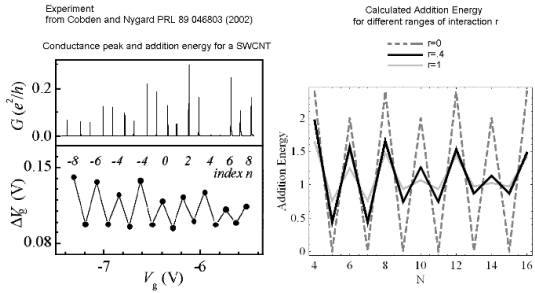}
 \caption{{On the right we show analytical Aufbau results
for the addition energy versus the number of electrons  of a
4-fold degeneracy model corresponding to different values of the
range $r$ ($r=0$ dark gray dashed line, $r=0.4$ black line, $r=1$
gray line). We show how the damping in the oscillations is due to
a long range interaction while it does not appear for a $r=0$
model. Our predictions can be compared with the measured addition energy in the Cobden
Nygard experiment displayed on the left.
}}
\end{figure}

\subsection{Asymmetric model}

Now we have to analyze models without  symmetries by using
eq.(\ref{ep9}) in the  HF approximation.

In order to introduce the electron electron interaction, we
take in account just two ($g^{\|}$ and $g^{\bot}$) of the many
constants that we introduced in section II from
eq.(\ref{intk}). For allowing to compare easily our results with those
in ref.\cite{Oreg}, as well as with experiments, we give our
results in terms of $V_0$ and $J$, obtained as a linear
combination of the  $g$ constants.

 Following the usual method,
in order to calculate the energy levels, we put
$$
  g^{\|}=V_0-J  \;  ; \quad g^{\bot}=V_0 \; ; \quad \Delta=\frac{v_F h}{2L}, \; ; \quad \Delta
\varepsilon=\delta N \frac{v_F h}{L}.
$$
Here $\Delta \varepsilon$ is the incommensurability shift
($\Delta \varepsilon<0.5 \Delta$).
The single particle energies have a different structure for
$\delta_{SM} >2\Delta \varepsilon$ and $\delta_{SM} <2\Delta
\varepsilon$, as we show in Fig.(1.d).

From the experimental data\cite{liang} we obtain
$$
V_0\approx U+\delta U +J_{exp}\approx .42 \quad and \quad J\approx
J_{exp}-2\delta U\approx .05
$$
where we assume $U=.22$, $\delta U=.05$ and $J_{exp}=.15$ in units of
$\Delta$. Under these conditions $J$ is always less than the level
spacing.

Now we can write the Hamiltonian of the nanotube depending on
these parameters\cite{Oreg}
\begin{eqnarray}\label{ep10}
 H &=&\sum_{{n,\zeta,p,s}}\varepsilon_{n,\zeta,p}
\hat{n}_{n,\zeta,p,s}  \\ \nonumber &+& V_0 \frac{N(N+1)}{2} -
J\sum_{{n,\zeta,p,s}}\sum_{{n`,\zeta`,p`,s}}\delta_{s,s`}\hat{n}_{n,\zeta,p,s}\hat{n}_{n,\zeta,p,s}
\end{eqnarray}

Since $J$ is less than the level spacing the energy that we need,
in order to add one electron is $\varepsilon$, corresponding to the
lowest empty energy level with an interaction energy
\begin{eqnarray}
V_0\frac{N(N-1)}{2} - J\frac{N-1}{2} \quad for \quad odd \quad N
\nonumber \\ V_0\frac{N(N-1)}{2} - J\frac{N}{2} \quad for \quad
even \quad N.\nonumber
\end{eqnarray}
Starting from  the Hamiltonian eq.(\ref{ep10}) we are able to
calculate the ground states of the many electron system for
various $N$.

Our results can  be compared to the experimental
results where strong asymmetries were found.
In Fig.(3) we show the peaks corresponding to
an asymmetric model to which we  apply the
classical theory of CB, i.e. eq.(\ref{gt}).
The fine structure with $8$ periodicity is destroyed by  thermal effects:
it is appreciable at smaller temperatures (we assume about $T_s\approx300\div500
mK$ for a Nanotube's length of about $100\div300 nm$) and disappears at a temperature $T\approx 4 T_s$ ($T\approx1.2\div2.0 K$ see ref.\cite{liang}) where just a $4$ periodicity appears. So, we conclude that we should not
be able to observe any small asymmetries effects if the temperature increases.
\begin{figure}
\includegraphics*[width=0.90\linewidth]{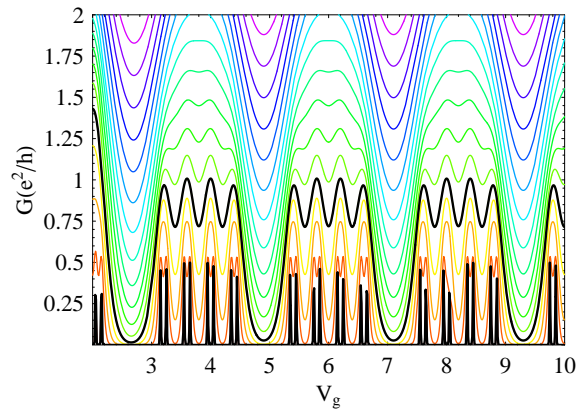}
 \caption{{The asymmetric model calculation for the
COs (conductance vs. gate voltage) at different
temperatures expressed in terms of $\Delta$ calculated following the classical CB theory. Theoretical
calculations show that  the fine structure  peaks are appreciable just for very low temperatures.
}}
\end{figure}
Fig.(3) shows how the Coulomb peaks in the conductance disappear
when the temperature increases. In a future article we will show
how the end of the CB regime corresponds to the beginning of
another one.

Now we want point out the limits of the approach used in the
previous sections: the HF approximation ignores the
effects of correlation. This corresponds to the Fermi Liquid
theory and gives good results just if we can assume that the
interaction is smaller than the kinetic energy ($g\ll\hbar v_F$).

Thermal effects are quite important too.  In fact the
temperature appears in the Beennaker formula and is responsible of
the disappearance of the Coulomb peaks. However, if the
temperature is higher for a SWCNT, it is the same Beennaker
formula which fails, because the HF calculated energy levels are
very different from the real energy levels of the electron system.
When $T$ is above a critical value
\begin{equation}\label{Tc}
    T_c= \frac{v_F  h}{L k_B}
\end{equation}
we cannot consider the Fermi sea as the ground state of the
electron system in a SWCNT because some other states with the same
linear momentum $K=0$ and different kinetic energy are also
available for the system. So for temperatures above the critical
value we have to take into account strong effects of correlation.

\subsection{Quantum Dots and Long Range Interactions}

As we did for a SWCNT  we  calculate the addition energy of a QD
in the HF approximation.

As we discussed in section II peaks and oscillations in the measured addition energy correspond to
the ones of a shell structure for a two-dimensional
harmonic potential. However,
in  experiments\cite{Tar},  high values of the addition
energy are observed also for $N=4,9,16,etc.$ corresponding to those values of the Number of
electrons in the dot for which, respectively, the second, third and fourth
shells are half filled with parallel spins in accordance with
Hund's rule. Half filled shells correspond to a maximum spin
state, which has a relatively low energy\cite{Tar,KeM}.

We compare a non-interacting model with models that include
Coulomb interactions especially the exchange term. In  Fig.(4) we
plot the calculated addition energy as a function of the number of electrons
in the following three different cases:

$\diamond$ The non-interacting model  gives us just the peaks
corresponding to the Magic Numbers.

$\diamond$ The model with constant parameters ($g^\|$ and
$g^\bot$) corresponds to a short range interaction (Dirac $\delta$).
The Coulomb exchange term in HF gives the  Hund's rule and
allows us to explain the oscillations in the addition energy as we
show in Fig.(4.a).

$\diamond$ The third model is the long range interaction one,
where we introduce two measured effects that are both due to the
long range Coulomb interaction. The first one  is due to the
classical capacitive effect shown in Fig.(4.b and c)  as a
continuous line. The second one
 is due to the long range correction to
Coulomb exchange.

We could compare these results to the experiments (e.g. see ref\cite{Tar}).
\begin{figure}
\includegraphics*[width=0.90\linewidth]{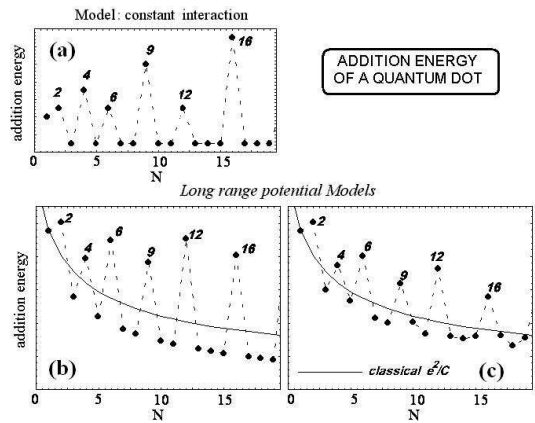}
 \caption{{In a),b),c) results from three
different models of interacting electrons in QDs are displayed. The
simple HF calculation in a) has to be corrected because it does
not take into account the classical capacitive effect.  In order to
determine it we recall that the electrons in the dot give a
charge droplet of  radius $R_D$ not fixed, so that we can
approximate it with a disk of capacitance $C= C_0 R_D$. The value of
$R_D$ can be calculated from  classical equations,
$R_D\propto \sqrt{N+1}$. So  we add the classical term to the
damped oscillations due to long range affected Coulomb exchange
and obtain the b) and c) plot in the figure.
}}
\end{figure}

\section{Conclusions}

In this work we have analyzed some properties of a 1D (SWCNT) and 0D-2D (QDs) electron systems
by introducing a model of interaction capable of interpolating from short to long range,
in order to analyze the effects of the long range component of electron electron interactions.

We have obtained that a damping in the oscillations of the
addition energy could be predicted for these models corresponding
to the presence of a long range interaction. The results can be
compared to the experiments and we discussed  the limits for the
experimental observation of our predictions  in SWCNT. In fact the
breakdown of the Fermi Liquid opens different regimes where the
tunneling has to be considered as resonant tunneling in
LLs\cite{812,NG}. So we have  to limit ourselves to describing an
electron system only when it is uncorrelated. This is true only if
the interaction strength is small ($g<<v_F h/L$)\cite{nota} and
the temperature is below the critical value  $T_c$ (see
eq.\ref{Tc}).

\acknowledgments

 \noindent This work was partially supported by the Italian Research Ministry MIUR, National Interest Program, under grant COFIN 2002022534.
\appendix

\section{Aufbau and energies for the 4-fold degenerate model}

We start by taking into account that each electron in the $|k|>k_F$
state interacts with all the electrons in the filled shells with
$k$ below $k_F$ so that we can have the following two (or four) $\Sigma_{HF}$
terms:
$$
\Sigma_{s,\sigma}(k_F,k)= \sum^{k_F}_{p=1}
U^{s,\sigma}_{|k-p|}+\sum^{-k_F}_{p=-1} U^{s,\sigma}_{|k-p|}
$$
If we introduce $n_F=\frac{L}{\pi}k_F$ and consider the direct
term of the interaction, we conclude that

\begin{eqnarray}
\nonumber   \Sigma_{\uparrow\downarrow}(k_F,k)&=& 2 V_0 n_F \\
\nonumber  \Sigma_{\uparrow\uparrow}(k_F,k)&=& V_0 \left(2 n_F  -
\sum^{k_F}_{p=-k_F}J_{|p-k|}(1-\delta_{p,0})\right)
\end{eqnarray}

Obviously if we add an electron in the lowest energy empty shell
we obtain $\Sigma(k_F,k_F+\frac{\pi}{L})\approx V_0 \left(4 n_F  -
\sum^{2 k_F+\frac{\pi}{L}}_{p=\frac{\pi}{L}}J_{p}\right)$.

For each shell we can consider the internal interaction energy as
$$
w(k)=2 V_0 (3-J_{2k})\qquad W(k_F)=\sum^{k_F}_{p=\frac{\pi}{L}}
w(p)
$$
so that the total energy of a system with $n_F$ filled shells reads
$$
E_{4n_F}=2 v_F n_F(n_F+1)+ W(k_F)+ 4
\sum^{k_F-\frac{\pi}{L}}_{p=0}   \Sigma(p,p+\frac{\pi}{L})
$$

\subsubsection{ Energy levels}

When we add one by one  the electrons to the $4 n_F$-system we
obtain
\begin{eqnarray}
 \nonumber \mu_{4n_F+1}&=&\hbar
v_F\left(k_F+\frac{\pi}{L}\right)+\Sigma(k_F,k_F+\frac{\pi}{L})
\\ \nonumber \mu_{4n_F+2}&=&\hbar
v_F\left(k_F+\frac{\pi}{L}\right)+\Sigma(k_F,k_F+\frac{\pi}{L})
+U_0(1-\gamma)
\\ \nonumber
\mu_{4n_F+3}&=&\hbar
v_F\left(k_F+\frac{\pi}{L}\right)+\Sigma(k_F,k_F+\frac{\pi}{L})
+2U_0 \\ \nonumber \mu_{4n_F+4}&=&\hbar
v_F\left(k_F+\frac{\pi}{L}\right)+\Sigma(k_F,k_F+\frac{\pi}{L})+U_0(3-\gamma)
\\ \nonumber
\mu_{4n_F+5}&=&\hbar v_F\left(k_F+2
\frac{\pi}{L}\right)+\Sigma(k_F+\frac{\pi}{L},k_F+2 \frac{\pi}{L})
\end{eqnarray}
where $\mu_{N+1}=E_{N+1}-E_{N}$ and $\gamma=\frac{J_{2 k_F}}{U_0}$

The addition energy is also calculated as
$E^A_{N+1}=\mu_{N+2}-\mu_{N+1}$ so that
\begin{eqnarray}
\nonumber E^A_{4n_F+1}&=& U_0(1-\gamma) \qquad
E^A_{4n_F+2}=U_0(1+\gamma)
\\ \nonumber E^A_{4n_F+3}&=&U_0(1-\gamma) \qquad
E^A_{4n_F+4}=U_0(1+\gamma) +\left(\frac{\hbar v_F \pi}{L}\right)
\end{eqnarray}


\bibliographystyle{prsty} 
\bibliography{}

\begin{thebibliography}{99}
\bibitem{nota}
This also corresponds to a long Nanotube so we cannot apply it to the experiments in ref\cite{postma} which concern short Nanotubes.
\end{thebibliography}

\end{document}